\documentclass[amsfonts,amssymb,amsmath,graphicx,color,latexsym,url,12pt]{article}
\usepackage{amsfonts,amssymb,amsmath}

\usepackage{caption}
\usepackage{subcaption}

\usepackage{xcolor}
\usepackage{mathrsfs} 
\usepackage{graphicx,type1cm,eso-pic,color}

\newcommand{\dsp}{\displaystyle}

\newcommand{\R}{{\mathbb R}}
\newcommand{\e}{{\mathrm e}}

\title{\bf Propagation of   Dirac spherical waves in  the expanding universe 
}

\author{{\bf Karen Yagdjian}
 }

\begin{document}

\date{}
\maketitle
\thispagestyle{empty}
\vspace{-0.3cm}

\begin{center}
{\it School of Mathematical and Statistical Sciences,
University of Texas RGV,\\
1201 W.~University Drive,  
Edinburg, TX 78539,
USA \\
E-mail: karen.yagdjian@utrgv.edu}
\end{center}
\medskip

\setcounter{equation}{0}
\pagenumbering{arabic}
\setcounter{page}{1}
\thispagestyle{empty}

\hspace{2cm}\begin{abstract}
\begin{small}
The explicit formulas for the spherical solutions of the Dirac equation in  the expanding universe are given. The   initial value of the solution can    be, in particular,  a wave function of the hydrogen-like atom or a spherical wave in the Minkowski space, that then propagates in the Friedmann-Lema\^itre-Robertson-Walker space-time, which is expanding with the de~Sitter  scale factor.  
\medskip

\end{small}
\end{abstract}

\section{Introduction} 

\setcounter{equation}{0}

The explicit formulas for spherical solutions of the Dirac equation in the Minkowski space are well known for a long time; see, e.g., \cite{Darwin,Bethe,Akhiezer,Greiner,Thaller}. Latest astronomical observational data confirm the accelerating expansion of the universe and stir up interest in spherical solutions in the space-times of such a universe. Especially intriguing and still unsolved is the problem of the wave function and the spectrum of the hydrogen atom in Friedmann-Lema\^itre-Robertson-Walker (FLRW) space-time, which is expanding with the exponential  scale factor. In this paper, we present   explicit formulas  for  the spherical solutions of the Dirac equation 
in the FLRW space-time with the de~Sitter  scale factor.

The diagonal metric tensor $g$ in the spatially flat  FLRW   space-time has the  entries 
$
g_{0 0 }=1$, $\,\,g_{1 1 }=g_{2 2 }=g_{3 3 }=-a^2(t)
$. 
We will focus on the space-time with the de~Sitter scale factor   $a(t)=e^{Ht}  $ (see, e.g., \cite{Moller}) that is modeling the expanding or contracting universe if $H>0$ or $H<0$, respectively. The Dirac equation in this space-time is (see, e.g., \cite{Barut-D})
\begin{equation}
\label{DE}
  \dsp 
\left(  i {\gamma }^0    \partial_0   +i e^{-Ht}{\gamma }^1  \partial_1+i  e^{-Ht}{\gamma }^2 \partial_2+i e^{-Ht}{\gamma }^3   \partial_3 +i \frac{3}{2}    H {\gamma }^0     -m{\mathbb I}_4 \right)\psi=0 \,,
\end{equation}
where 
the contravariant gamma matrices are   (see,  e.g., \cite[p. 61]{B-Sh})
\begin{equation}
 \gamma ^0= \left (
   \begin{array}{ccccc}
   {\mathbb I}_2& {\mathbb O}_2   \\
   {\mathbb O}_2& -{\mathbb I}_2   \\ 
   \end{array}
   \right),\quad 
\gamma ^k= \left (
   \begin{array}{ccccc}
  {\mathbb O}_2& \sigma ^k   \\
  -\sigma ^k &  {\mathbb O}_2  \\  
   \end{array}
   \right),\quad k=1,2,3\,.
 \end{equation}
Here $\sigma ^1 $, $\sigma ^2 $, ans $\sigma ^2 $ are Pauli matrices (see, e.g.,  \cite[p.97]{B-L-P}), 
while  ${\mathbb I}_n $, ${\mathbb O}_n $ denote the $n\times n$ identity and zero matrices, respectively.  

The problem of the hydrogen atom in an expanding universe has interested researchers  since time when  quantum mechanics provided an exhaustive answer to that problem in the Minkowski space   \cite{Darwin}. The equation for the hydrogen atom in an exponentially expanding universe  is
\begin{equation}
\label{HydrogenDS} 
\left(  i {\gamma }^0    \partial_0   +i \e^{-Ht}{\gamma }^\ell  \partial_\ell   +i     \frac{3}{2}   H  {\gamma }^0
-\frac{e}{\hbar c}{\gamma }^0 V  -m{\mathbb I}_4 \right) \Psi   
  =  0.   
\end{equation}
From now on, the   Einstein summation convention is used. Here $V=V(r)=const  \cdot r^{-1}$ and the solution $\Psi $ must have the finite $L^2$ norm, 
$
\int_{{\mathbb R}^3}\|\Psi (x,t)\|^2\,dx <\infty$.    
The complexity of finding an explicit formula for solutions to  equation (\ref{HydrogenDS}) is indicated by the numerous publications that present diversified variations of the problem, including changes in the field equation, for example, replacing the Dirac equation with the Klein-Gordon equation \cite{Veko} or with Newtonian analysis of ``classical atom'' \cite{Price-Romano}, or by requiring the Coulomb potential to decay in time as $e^{- Ht}$ for the Dirac field and $e^{- 2Ht}$ for the scalar field (see \cite{Baloi} and references therein). The simultaneous presence of the electrical field and cosmological expansion, that is, the functions $V(r)=const \cdot r^{-1}$ and $e^{Ht}$, respectively, causes this complexity. The cases where there is only a single complication, either the potential $V(r)$ or the scale factor $e^{Ht}$ (with $H\not= 0$), have been well studied; see \cite{Bethe, Baskin} and \cite{AnPH2020}, respectively.

In this paper, we unify these complexities by prescribing specific initial data $\Psi(x,0)$ generated by the potential, thereby bringing the electric field into the problem. Thus, we allow evidence of the electrical field only in the initial state (spinor), while the cosmological constant (expansion) is present in the equation of the field propagating in the expanding background. We choose the spherical initial state, with the hydrogen atom in particular in mind. One can regard our approach as a first step toward approximating the solution to the difficult problem of developing the spectral nature of the initial state under evolution in the expanding universe.

Nowotny \cite{Nowotny} studied a hydrogen atom embedded in a spatially isotropic and homogeneous expanding space-time by considering the Dirac equation. He provided an exact analysis of the structure of the energy eigenvalue spectrum in the static FLRW universe. In \cite{Nowotny}, for the case of hyperbolic space, it is proved that a discrete energy spectrum exists in the same energy range as for flat space. In contrast, for the case of spherical space, the energy spectrum is continuous in the entire energy range.

In the case of bounded electrons, McVittie \cite{McVittie} analyzed the energy spectrum of the hydrogen atom based on the  Dirac equation using a perturbation method. In the case of an expanding, closed, or open universe, he found that the energy levels of the hydrogen atom are only slightly influenced by the expansion or contraction of the universe. Dautcourt \cite{Dautcourt} has studied the impact of the structure of space-time on the energy spectrum of the hydrogen atom by a perturbational approach that starts with the well-known results for   flat space. In the case of the static Einstein universe, he found a discrete energy spectrum that differs from that of the hydrogen atom in flat space.

 De~Oliveira and Schmidt \cite{deOliveira-Schmidt} studied the Dirac equation in a  static curved space--time, where the line element of the metric   $ds^2 = e^{2f (r)}dt^2 - e^{2g(r)}dr^2 -r^2d\theta^2 - r^2 \sin^2\theta d\phi^2$ depended on two arbitrary functions $f (r)$ and $g(r)$. In particular, for the case of a metric model $f = g$, they uncoupled the radial component of the spinor and found exact solutions for the hydrogen atom. 
They   discussed three target systems, namely, Schwarzschild, anti--de Sitter, and de Sitter space-times.

For the Dirac equation with potential, which   models the hydrogen atom, Baskin and Wunsch \cite{Baskin} showed that the singularities 
of the Schwartz kernel of the propagator are along an expanding spherical wave away from rays that miss the potential singularity at the 
origin, but also may include an additional spherical wave of diffracted singularities emanating from the origin. In particular, the diffracted wavefront is $1-\varepsilon $ 
derivatives smoother than the main singularities, for all $\varepsilon  > 0$, and it  is a conormal singularity.

The rest of this paper is organized as follows. In Section~\ref{S2}, we describe the explicit formulas for solutions to the Dirac equation in the expanding universe (Subsection~\ref{SS2.1}). Then we give the explicit formulas for the spherical wave in the expanded universe produced by the initial state, which is a free spherical wave in Minkowski space (Subsection~\ref{SS2.2}). In Subsections~\ref{SS2.3}, \ref{SS2.4}, and \ref{SS2.5} are given solutions for the waves of the massless particle and of the particles with imaginary mass $iH$ and $-iH$, respectively. The  last three waves obey Huygens' principle (see, \cite{WunschV,JPA2021}). Particles with imaginary mass are called ``tachyon''; since the  last century, tachyons have been continuously studied by numerous authors in various physical contexts  (see, e.g., \cite{Feinberg}-\cite{Moucherek}  
%,Ross,Chan,Moucherek,Schwartz,Schwartz2,Paczos,Jodlowski,Jodlowski2,}
and references therein).
For these three waves, we indicate the best decay rate in the $L_\infty$ space. We discover that the cosmological constant causes the damping of oscillations, which vanish after a certain life-span time $T_{ls.osc}$.  In Subsection~\ref{SS2.6} we discuss the wave produced by the hydrogen-like atom. The appendix contains a short list of necessary formulas in the Minkowski space.

\section{Spherical Dirac waves in expanding universe}
\label{S2}

The Dirac equation in de~Sitter space-time (\ref{DE}) 
  can be written in the spherical coordinates as follows
\begin{equation}
\label{DiracSpher}
 \left(i {\gamma }^0  \partial_0+i   \e^{-Ht} \left( {\gamma }^r_c \partial_r  +{\gamma }_c^\theta  \partial_\theta+ {\gamma }_c^\phi  \partial_\phi  \right) + i       \frac{3}{2}  H  {\gamma }^0    -M{\mathbb I}_4\right)\Psi =0 \,,
\end{equation}
where in this Cartesian tetrad gauge the gamma matrices will be given by (see, e.g., \cite{Schluter}) 
\begin{eqnarray}
{\gamma }^r_c
& = &
\gamma ^1 \cos (\phi ) \sin (\theta )
+\gamma ^2 \sin (\theta ) \sin (\phi ) 
+ \gamma ^3  \cos (\theta )\,,\\
%\label{3.15}
{\gamma }^\phi _c 
& = &  \frac{ 1}{r \sin (\theta  )}\left( -\gamma ^1\sin (\phi  ) 
+ \gamma ^2 \cos (\phi  ) \right)  \,,\\
%\label{3.16}
{\gamma }^\theta _c 
& = &
\frac{ 1 }{r}\left(\gamma ^1\cos (\theta ) \cos (\phi ) 
+ \gamma ^2\sin (\phi) \cos (\theta  ) 
-\gamma ^3 \sin (\theta  )    \right)\,.
\end{eqnarray}
We have used the subscript $c$ for Cartesian. 
Denote
$M_+=\frac{1}{2}H+iM$, $M_-=\frac{1}{2}H-iM$. 
According to Theorem~0.2~\cite{AnPH2020}, 
  the solution to the Cauchy problem
\begin{equation} 
\begin{cases} \dsp \left(i {\gamma }^0  \partial_0+i \e^{-Ht} \gamma ^k  \partial_k   + i       \frac{3}{2}  H  {\gamma }^0   -M{\mathbb I}_4\right)\Psi  =0 ,\cr 
\Psi (x,0)= \Phi   (x ) 
\end{cases}
\end{equation} 
can be represented  by the following formula
\begin{equation}
\label{MainRep} 
\hspace{-0.2cm} \Psi (x,t) 
 =  
i{\mathscr{D}}^{co}(t,\partial _t,\partial _x;M)  \left (
   \begin{array}{cccc}
 {\cal K}_1(x,t,D_x;M_+) {\mathbb I}_2&{\mathbb O}_2  \\
 {\mathbb O}_2 & -{\cal K}_1(x,t,D_x;M_-) {\mathbb I}_2 \\
   \end{array}
   \right ) \Phi  (x,t)  \,,
 \end{equation} 
 where the operator 
\begin{equation}
\label{Dco} 
{\mathscr{D}}^{co}(t,\partial _t,\partial _x;M) 
  :=  
- \e^{-Ht}\left( i\gamma ^0 \partial_0+  i\e^{-Ht}\left({\gamma }^r_c \partial_r  +{\gamma }_c^\theta  \partial_\theta+ {\gamma }_c^\phi  \partial_\phi  \right)- i\frac{H}{2}\gamma ^0+M{\mathbb I}_4\right),   
 \end{equation}
due to Proposition~3.4~\cite{AnPH2020}, is the right co-factor ${\mathscr{D}}^{co}(t,\partial _t,\partial _x;M)$ (the right diagonalizer of the Dirac operator (\ref{DiracSpher})), that is,
\begin{eqnarray} 
&   &
 \left(i {\gamma }^0  \partial_0+i   \e^{-Ht} \left({\gamma }^r_c \partial_r  +{\gamma }_c^\theta  \partial_\theta+ {\gamma }_c^\phi  \partial_\phi    \right)  
+ i       \frac{3}{2}  H  {\gamma }^0   -m{\mathbb I}_4\right) {\mathscr{D}}^{co}(t,\partial _t,\partial _x;m) \nonumber \\
& = &   
\e^{-Ht}\Big[\left(  \partial_0^2     
-  \e^{- 2Ht}    
\left(   \frac{ \partial^2 }{ \partial r^2}  
+\frac{2}{r } \frac{ \partial }{ \partial r} 
+ \frac{1}{r^2 } \frac{ \partial^2 }{ \partial \theta ^2 }    
+ \frac{\cot (\theta ) }{r^2 } \frac{ \partial }{ \partial \theta  }  
 + \frac{1}{r^2\sin^2 (\theta )} \frac{ \partial^2 }{ \partial \phi ^2  }  \right)\right){\mathbb I}_4  \nonumber \\   
&  &
+ \left( m{\mathbb I}_4  - \frac{1}{2}iH \gamma ^0\right)^2\Big]  \,.
\end{eqnarray} 
 Here the integral transform operator ${\cal K}_1(x,t,D_x;M)$    is defined (see \cite{Yag_Galst_CMP,JMP_2024}) as follows,
\begin{equation}
\label{KdxDef}
{\cal K}_1(x,t,D_x;M) \varphi (x) = 
 2\int_{0}^{\frac{1}{H}\left(1-\e^{-Ht}\right)} 
  K_1( s,t;M)  \int_{{\mathbb R}^3} {\mathcal E}^w (x-y,s)  \varphi  (y) \,d y\, ds\,, \quad \varphi \in C_0^\infty({\mathbb R}^n),
\end{equation}
with the  kernel  
\begin{eqnarray}
\label{K1}
K_1(r,t;M)
& :=  &
4^{-\frac{M}{H}} \e^{ M t} \left(\left(1+\e^{-H t}\right)^2-(H r)^2\right)^{\frac{M}{H}-\frac{1}{2}}  \\
&  &
\times  F \left(\frac{1}{2}-\frac{M}{H},\frac{1}{2}-\frac{M}{H};1;\frac{\left(1-\e^{-H t}\right)^2-(r H)^2}{\left(1+\e^{-H t}\right)^2-(r H)^2}\right) \nonumber\,.   
\end{eqnarray} 
The function   $F(a,b;c;\gamma) $ is   the   hypergeometric function (see, e.g., \cite{B-E}). 
The distribution ${\mathcal E}^w (x,s) $ is a  kernel of the solution operator to the problem
\begin{eqnarray}
\label{we} 
v_{tt}-  \bigtriangleup v =0, \quad v(x,0)= \varphi  (x)
, \quad v_t(x,0)= 0\,,
\end{eqnarray}
so that  the function $v(x,t)=\int_{{\mathbb R}^3} {\mathcal E}^w (x-y,s)  \varphi  (y) \,d y $  solves problem (\ref{we}). The Laplace  operator in ${\mathbb R}^3$ is denoted by $ \Delta$.

Thus, as a consequence of Theorem~0.2~\cite{AnPH2020} and the results of \cite{ArXiV2025},  all    Dirac waves and, in particular, spherical waves in the de~Sitter universe are explicitly represented by (\ref{MainRep}) through their value at some  moment of time. We are especially interested in the waves, which at the initial time are the wave functions of the hydrogen-like atom in the Minkowski space.

\subsection{Explicit representation for the spherical wave  in the expanding universe}
\label{SS2.1}

We begin by calculating the spinor 
 \begin{eqnarray}
 \label{K1diag}
&  &
\left (
   \begin{array}{cccc}
 {\cal K}_1(x,t,D_x;M_+){\mathbb I}_2 &{\mathbb O}_2  \\ 
{\mathbb O}_2&  {\cal K}_1(x,t,D_x;M_-){\mathbb I}_2   \\ 
   \end{array}
   \right ) \Phi   \,,
 \end{eqnarray} 
of the  formula (\ref{MainRep}), where the initial value $\Phi $ of the wave (spinor) is chosen either by (\ref{eelplas1/2}) and (\ref{A4}) with $t=0$, or by (\ref{eelminus1/2}) and (\ref{A5}) with $t=0$.

The functions $ f$ and $ g $ will be specified below in two cases: (1) for the   
waves generated by the initial state $\Phi$, which is a state of the free particle  in the Minkowski space (see, e.g., \cite[Sec.24]{B-L-P});   (2) for the   
waves generated by the initial state $\Phi$, which is a state of the hydrogen-like atom in the Minkowski space (see, e.g., \cite[Sec.14]{Bethe}).

The kernel function is defined by (\ref{K1}),  
while the scalar operator ${\cal K}_1(x,t,D_x;M)$ acts  by (\ref{KdxDef}).
The entries of (\ref{K1diag}) 
 can be written as follows
\begin{equation} 
{\cal K}_1(x,t,D_x;M)  F_0 (r)  Y_{\ell m}(\theta, \phi)  
 = 
 2Y_{\ell m}(\theta, \phi)\int_{0}^{\frac{1}{H}\left(1-\e^{-Ht}\right)} 
  K_1( s,t;M)   F (r,s)   \, ds
\end{equation}
with the corresponding $M_-$, $M_+$, and $F (r,s) $.  The   radial function $F (r,s) $ is a solution to the problem
\begin{equation} 
\begin{cases}
\dsp F_{tt}(r,t)-F_{rr}(r,t)
-    \frac{2}{r} F_r(r,t)+\frac{\ell(\ell+1)}{r^2}F(r,t)   =0\,,\\ 
F (r,0)=F_0 (r),\quad F _t(r,0)=0\,,
\end{cases} 
\end{equation}
where the function $F_0 (r)$ will be chosen later, and     
\begin{eqnarray}
\label{aroundzeroR}
\hspace{-0.9cm} &  &
F_0(r)= 
O(r^{\mu -\frac{1}{2}}) \,\, \mbox{\rm as } \,\, r \searrow 0 \,,
\end{eqnarray}
with the parameter $\mu$, which will be specified below. Thus,
\begin{equation}
{\cal K}_1(x,t,D_x;M)  F_0 (r)  Y_{\ell m}(\theta, \phi) 
 = 
Y_{\ell m}(\theta, \phi)  \widetilde{F} (r,t) \,,
\end{equation}
where, according to (\ref{K1}), we have denoted
\begin{eqnarray}
\widetilde{F} (r,t)
& := &
 4^{\frac{1}{2}-\frac{M}{H}}  \e^{M t}\int_{0}^{\frac{1}{H}\left(1-\e^{-Ht}\right)} 
 \Bigg\{  \left(\left(1+\e^{-H t}\right)^2-(H s)^2\right)^{\frac{M}{H}-\frac{1}{2}} \\
&  &
\times  F \left(\frac{1}{2}-\frac{M}{H},\frac{1}{2}-\frac{M}{H};1;\frac{\left(1-\e^{-H t} \right)^2-(s H)^2}{\left(1+\e^{-H t} \right)^2-(s H)^2}\right) \Bigg\}      F (r,s)   \, ds\,. \nonumber
\end{eqnarray}

\subsection{Solution with an initial wave produced by a free spherical wave}
\label{SS2.2}

For the free wave in the expanding universe,  which is produced by a free spherical wave, we choose the initial state, according to (\ref{MainRep}) and Section~\ref{ss4} (see  (\ref{eelplas1/2}),(\ref{A4})) and derive
\begin{eqnarray} 
\Psi (x,t) 
& = &
{\mathscr{D}}^{co}(t,\partial _t,\partial _x;M) i\gamma ^0 \left(
\begin{array}{c}
 {\cal K}_1(x,t,D_x;M_+) i f(r) \sqrt{\frac{\ell -m+\frac{3}{2}}{2 \ell +3}}  Y_{\ell+1}^{m-\frac{1}{2}}(\theta ,\phi ) \\
  {\cal K}_1(x,t,D_x;M_+) i f(r) \sqrt{\frac{\ell +m+\frac{3}{2}}{2 \ell +3}}Y_{\ell+1}^{m+\frac{1}{2}}(\theta ,\phi )  \\
 {\cal K}_1(x,t,D_x;M_-)g(r) \sqrt{\frac{\ell +m+\frac{1}{2}}{2 \ell +1}}  Y_{\ell }^{m-\frac{1}{2}}(\theta ,\phi ) \\
- {\cal K}_1(x,t,D_x;M_-)g(r) \sqrt{\frac{\ell -m+\frac{1}{2}}{2 \ell +1}}  Y_{\ell }^{m+\frac{1}{2}}(\theta ,\phi )  \\
\end{array}
\right) \,.
 \end{eqnarray}
For  $M_+=\frac{1}{2}H+iM $, we obtain
\begin{equation}
 {\cal K}_1(x,t,D_x;M_+) i f(r)  Y_{\ell+1}^{m-\frac{1}{2}}(\theta ,\phi )\\
=  
iY_{\ell+1}^{m-\frac{1}{2}}(\theta ,\phi )\widetilde{F_{\ell +1}}  (r,t)
\,,
\end{equation}
 where
\begin{eqnarray}
\label{Ftilde}
\widetilde{F_{\ell +1}}  (r,t)  
& = & 
4^{    -i\frac{M}{H} }  \e^{ t(\frac{1}{2}H+iM)} 
 \int_{0}^{\frac{1}{H}\left(1-\e^{-Ht}\right)} 
 \Bigg\{  \left(\left(1+\e^{-H t}\right)^2-(H s)^2\right)^{i\frac{M}{H}} \\
&  &
\times  F \left( -i\frac{M }{H},-i\frac{M }{H};1;\frac{\left(1-\e^{-H t} \right)^2-(s H)^2}{\left(1+\e^{-H t} \right)^2-(s H)^2}\right) \Bigg\}      F_{\ell +1} (r,s)   \, ds  \,.\nonumber
 \end{eqnarray}
The function $F_{\ell +1} (r,s) $ is a solution to the problem
\begin{eqnarray}
\label{Fell}
\begin{cases}
\dsp F_{tt}(r,t)-F_{rr}(r,t)
-    \frac{2}{r} F_r(r,t)+\frac{(\ell+1)(\ell+2)}{r^2}F(r,t)   =0\,,\\ 
F (r,0)=f (r) ,\quad F _t(r,0)=0\,.
\end{cases}
 \end{eqnarray}
Similarly,   
 \begin{equation}
{\cal K}_1(x,t,D_x;M_+) i f(r) Y_{\ell+1}^{m+\frac{1}{2}}(\theta ,\phi )   =
iY_{\ell+1}^{m+\frac{1}{2}}(\theta ,\phi )\widetilde{F_{\ell +1}}  (r,t) \,. 
 \end{equation}
For the second pair of the  entries with   $M_-=\frac{1}{2}H-iM $,  we obtain
\begin{equation}
{\cal K}_1(x,t,D_x;M_-)g(r)   Y_{\ell }^{m-\frac{1}{2}}(\theta ,\phi ) =
 Y_{\ell }^{m-\frac{1}{2}}(\theta ,\phi )  \widetilde{G_\ell }(r,t)\,, 
\end{equation}
where    
 \begin{eqnarray} 
\label{Gtilde}
 \widetilde{G_\ell} (r,t)
& = &
  4^{  i\frac{M}{H} }  \e^{ t(\frac{1}{2}H-iM)}\int_{0}^{\frac{1}{H}\left(1-\e^{-Ht}\right)} 
 \Bigg\{  \left(\left(1+\e^{-H t}\right)^2-(H s)^2\right)^{-i\frac{M}{H}} \\
&  &
\times  F \left(i\frac{M}{H},i\frac{M}{H};1;\frac{\left(1-\e^{-H t} \right)^2-(s H)^2}{\left(1+\e^{-H t} \right)^2-(s H)^2}\right) \Bigg\}    G_\ell (r,s)   \, ds  \,. \nonumber
\end{eqnarray}
The function $G_\ell (r,s) $ is a solution to the problem
\begin{eqnarray}
\label{Gell}
\begin{cases}
\dsp G_{tt}(r,t)-G_{rr}(r,t)
-    \frac{2}{r} G_r(r,t)+\frac{\ell(\ell+1)}{r^2}G(r,t)   =0\,,\\ 
G (r,t)=g (r),\quad G_t (r,0)=0\,.
\end{cases}
 \end{eqnarray}
In this case we derive
\begin{equation} 
- {\cal K}_1(x,t,D_x;M_-)g(r)   Y_{\ell }^{m+\frac{1}{2}}(\theta ,\phi ) 
   =  
  -Y_{\ell }^{m+\frac{1}{2}}(\theta ,\phi ) \widetilde{G_\ell} (r,t)\,.
\end{equation}
Thus, for $ j=\ell+\frac{1}{2}$ we obtain
\begin{equation}
\left(
\begin{array}{c}
 {\cal K}_1(x,t,D_x;M_+) i f(r) \sqrt{\frac{\ell -m+\frac{3}{2}}{2 \ell +3}}  Y_{\ell+1}^{m-\frac{1}{2}}(\theta ,\phi ) \\
  {\cal K}_1(x,t,D_x;M_+) i f(r) \sqrt{\frac{\ell +m+\frac{3}{2}}{2 \ell +3}}Y_{\ell+1}^{m+\frac{1}{2}}(\theta ,\phi )  \\
 {\cal K}_1(x,t,D_x;M_-)g(r) \sqrt{\frac{\ell +m+\frac{1}{2}}{2 \ell +1}}  Y_{\ell }^{m-\frac{1}{2}}(\theta ,\phi ) \\
- {\cal K}_1(x,t,D_x;M_-)g(r) \sqrt{\frac{\ell -m+\frac{1}{2}}{2 \ell +1}}  Y_{\ell }^{m+\frac{1}{2}}(\theta ,\phi )  \\
\end{array}
\right) 
=
\left(
\begin{array}{c}
i\sqrt{\frac{\ell -m+\frac{3}{2}}{2 \ell +3}}Y_{\ell+1}^{m-\frac{1}{2}}(\theta ,\phi )\widetilde{F_{\ell +1}}  (r,t) \\
 i\sqrt{\frac{\ell +m+\frac{3}{2}}{2 \ell +3}}Y_{\ell+1}^{m+\frac{1}{2}}(\theta ,\phi )\widetilde{F_{\ell +1}}  (r,t)  \\
\sqrt{\frac{\ell +m+\frac{1}{2}}{2 \ell +1}} Y_{\ell }^{m-\frac{1}{2}}(\theta ,\phi )  \widetilde{G_\ell} (r,t)\\
 -\sqrt{\frac{\ell -m+\frac{1}{2}}{2 \ell +1}}Y_{\ell }^{m+\frac{1}{2}}(\theta ,\phi ) \widetilde{G_\ell} (r,t)  \\
\end{array}
\right) ,
 \end{equation} 
 while for $j=\ell-\frac{1}{2} $  we have
\begin{equation}
\left(
\begin{array}{c}
 {\cal K}_1(x,t,D_x;M_+) i f(r) \sqrt{\frac{\ell +m-\frac{1}{2}}{2 \ell-1}}  Y_{\ell-1}^{m-\frac{1}{2}}(\theta ,\phi ) \\
 - {\cal K}_1(x,t,D_x;M_+) i f(r) \sqrt{\frac{\ell -m-\frac{1}{2}}{2 \ell -1}}Y_{\ell-1}^{m+\frac{1}{2}}(\theta ,\phi )  \\
 {\cal K}_1(x,t,D_x;M_-)g(r) \sqrt{\frac{\ell -m+\frac{1}{2}}{2 \ell +1}}  Y_{\ell }^{m-\frac{1}{2}}(\theta ,\phi ) \\
 {\cal K}_1(x,t,D_x;M_-)g(r) \sqrt{\frac{\ell +m+\frac{1}{2}}{2 \ell +1}}  Y_{\ell }^{m+\frac{1}{2}}(\theta ,\phi )  \\
\end{array}
\right) 
=
\left(
\begin{array}{c}
 i   \sqrt{\frac{\ell +m-\frac{1}{2}}{2 \ell-1}}  Y_{\ell-1}^{m-\frac{1}{2}}(\theta ,\phi )\widetilde{F_{\ell -1}}(r,t) \\
 -  i  \sqrt{\frac{\ell -m-\frac{1}{2}}{2 \ell -1}}Y_{\ell-1}^{m+\frac{1}{2}}(\theta ,\phi )\widetilde{F_{\ell -1}}(r,t)  \\
  \sqrt{\frac{\ell -m+\frac{1}{2}}{2 \ell +1}}  Y_{\ell }^{m-\frac{1}{2}}(\theta ,\phi )\widetilde{G_\ell }(r,t) \\
  \sqrt{\frac{\ell +m+\frac{1}{2}}{2 \ell +1}}  Y_{\ell }^{m+\frac{1}{2}}(\theta ,\phi ) \widetilde{G_\ell }(r,t) \\
\end{array}
\right) . 
\end{equation}
 
The functions $F_{\ell +1} (r,s) $ and $G_\ell(r,s) $, which solve problems (\ref{Fell}) and (\ref{Gell}), respectively,  can be written in three different ways (see \cite[Sec.2]{ArXiV2025}):  via the general solution of the equation (\ref{Fell}), 
via the Riemann function, or, if certain conditions (\ref{intF}) are fulfilled (see \cite{ArXiV2025}), via the Hankel transform  that is more amenable for the $L^2({\mathbb R}^3)$-estimates. The way of writing depends on the initial functions $f(r)$ and $g(r)$. For the initial data produced by a free wave, we must use the Riemann function. For $\ell=0,1,2,3,4$, a  simple form for these functions is given in \cite{ArXiV2025}. 
  
 Assume  that $ f,g \in C^\infty(\R_+)$, condition (\ref{aroundzeroR}) is fulfilled,  and that $\mu  >\ell-\frac{3}{2}$. The functions $f(r)$ and $g(r)$ will be continued to negative $r$ as odd functions for odd $\ell$ and as even functions for even $\ell$, that is, $f(-r)=(-1)^\ell f(r)$ and $g(-r)=(-1)^\ell g(r)$.  Then according to Lemma~2.2~\cite{ArXiV2025}, the functions $F_\ell (r ,t ) $ and $ G_\ell (r ,t ) $ are given by  
\begin{eqnarray}  
\label{4.1}
F_{\ell+1} (r ,t )  
& = &
  \frac{1}{2r } \Big[  (r-t )f(r-t )    + (r+t) f(r+t) \Big]
\\
&  &
-\frac{1}{4 } (\ell +1)  (\ell +2) \frac{t}{r^2}\int_{r-t}^{r+t} f(s)  
  F \left(-\ell ,\ell +3;2;\frac{t^2-(r-s )^2}{4 r s }\right)ds\,, \nonumber \\
\label{4.2}
G_\ell (r ,t )  
& = &
  \frac{1}{2r } \Big[  (r-t )g(r-t )    + (r+t) g(r+t) \Big]\\
  &  &
-\frac{1}{4 } \ell  (\ell +1) \frac{t}{r^2}\int_{r-t}^{r+t} g(s)  
  F \left(1-\ell ,\ell +2;2;\frac{t^2-(r-s )^2}{4 r s }\right)ds\,.\nonumber 
\end{eqnarray}
Note: for a positive integer $\ell\geq 1$, the functions $F \left(-\ell ,\ell +3;2;z\right)$ and $F \left(1-\ell ,\ell +2;2;z\right) $ are   polynomials.  Then,  we obtain
\begin{equation} 
\Psi (x,t) 
  = 
 -{\mathscr{D}}^{co}(t,\partial _t,\partial _x) i\gamma ^0\left(
\begin{array}{c}
  \sqrt{\frac{\ell -m+\frac{3}{2}}{2 \ell +3}}Y_{\ell+1}^{m-\frac{1}{2}}(\theta ,\phi )\widetilde{F_{\ell +1}}  (r,t) \\
 \sqrt{\frac{\ell +m+\frac{3}{2}}{2 \ell +3}}Y_{\ell+1}^{m+\frac{1}{2}}(\theta ,\phi )\widetilde{F_{\ell +1}}  (r,t)  \\
  i\sqrt{\frac{\ell +m+\frac{1}{2}}{2 \ell +1}} Y_{\ell }^{m-\frac{1}{2}}(\theta ,\phi )  \widetilde{G_\ell} (r,t)\\
-i\sqrt{\frac{\ell -m+\frac{1}{2}}{2 \ell +1}}Y_{\ell }^{m+\frac{1}{2}}(\theta ,\phi ) \widetilde{G_\ell} (r,t)  \\
\end{array}
\right) \,\, \mbox{\rm for } \,\, j=\ell+\frac{1}{2}\,. 
\end{equation}
Note the functions $\widetilde{F_{\ell +1}}  (r,t)$ and $ \widetilde{G_\ell} (r,t) $ satisfy  the folowing   Klein-Gordon equations with the ``imaginary mass'' $M_\pm $ in the de~Sitter spacetime  (see Proposition 3.4~\cite{AnPH2020})
\begin{eqnarray} 
&  &    
\left(  \partial_0^2     
-  \e^{- 2Ht}    
\left(   \frac{ \partial^2 }{ \partial r^2}  
+\frac{2}{r } \frac{ \partial }{ \partial r} 
- \frac{(\ell+1)(\ell+2)}{r^2 }  \right)\right)\widetilde{F_{\ell +1}}  (r,t) 
- M_+ ^2 \widetilde{F_{\ell +1}}  (r,t) =0,\\
&   &
\left(  \partial_0^2     
-  \e^{- 2Ht}    
\left(   \frac{ \partial^2 }{ \partial r^2}  
+\frac{2}{r } \frac{ \partial }{ \partial r} 
- \frac{\ell(\ell+1)}{r^2 }  \right)\right) \widetilde{G_\ell} (r,t) 
-  M _ - ^2 \widetilde{G_\ell} (r,t)=0 \,.
\end{eqnarray} 
respectively. (Compare with \cite[(14.5),(14.6)]{Bethe}.) Similarly, we obtain
\begin{equation}
\Psi (x,t) 
  =  
 - {\mathscr{D}}^{co}(t,\partial _t,\partial _x;M)  i\gamma ^0 \left(
\begin{array}{c}
     \sqrt{\frac{\ell +m-\frac{1}{2}}{2 \ell-1}}  Y_{\ell-1}^{m-\frac{1}{2}}(\theta ,\phi )\widetilde{F_{\ell -1}}(r,t) \\
- \sqrt{\frac{\ell -m-\frac{1}{2}}{2 \ell -1}}Y_{\ell-1}^{m+\frac{1}{2}}(\theta ,\phi )\widetilde{F_{\ell -1}}(r,t)  \\
    i \sqrt{\frac{\ell -m+\frac{1}{2}}{2 \ell +1}}  Y_{\ell }^{m-\frac{1}{2}}(\theta ,\phi )\widetilde{G_\ell} (r,t) \\
   i\sqrt{\frac{\ell +m+\frac{1}{2}}{2 \ell +1}}  Y_{\ell }^{m+\frac{1}{2}}(\theta ,\phi ) \widetilde{G_\ell} (r,t) \\
\end{array}
\right) \,\, \mbox{\rm for }\,\, j=\ell-\frac{1}{2}\,.
 \end{equation}

\subsection{Example of the massless particle with $\ell=0$}
\label{SS2.3}

To illustrate these formulas we consider the waves, which obey 
 the Huygens' principle in the de~Sitter space, that is, the waves of the spin-1/2 particles with $M=0$, $iH$, and $-iH$ (see, \cite{WunschV,JPA2021}). For the massless  wave  $M=0$, we choose the initial spinor as the one in Minkowski space with the value at $t=0$ and $\ell=0$ in (\ref{eelplas1/2}) and (\ref{A4}).
 For $ M=0$, we have
 $
 M_+=M_-=\frac{1}{2}H$, 
and 
\begin{eqnarray}
\label{K1M0}
K_1\left(r,t;\pm\frac{1}{2}H \right)
  :=  
\frac{1}{2} \e^{\frac{H t}{2}},\quad  
K_1\left(r,t;\frac{3}{2}H\right)
  :=  \frac{1}{4} \e^{-\frac{1}{2} (H t)} \left(\left(1-H^2 r^2\right) \e^{2 H t}+1\right)\,.  
\end{eqnarray}
Hence, according to (\ref{A4}), for  $M=0$, $\ell=0$,  and $j=\ell +\frac{1}{2}$,  we obtain the odd and even functions
\begin{eqnarray} 
f(r)
  =  
 c_1\frac{1}{p^2 r^2} (p r  \cos (p r )-\sin (p r )) ,\quad  
g(r) 
  = 
c_1\frac{ \sin (p r )}{ {p r }}   \,,
\end{eqnarray}
respectively. We set $c_1=1$ for simplicity. Then for these  $M$, $\ell$, and $ j$, according to (\ref{4.1}) and (\ref{4.2}), we derive
\begin{eqnarray}
F_1 (r,t)
  =     
\frac{1}{p^2 r^2} \Big(\cos (p t) (p r \cos (p r)-\sin (p r))\Big), 
\quad
G_0 (r,t)
  =  
\frac{1}{p r}\sin (p r) \cos (p t)\,. %\quad r \in {\mathbb R}, \,\,t\geq 0  .
\end{eqnarray}
The  functions $\widetilde{F_1} $ and $\widetilde{G_0} $  are
\begin{eqnarray}
\widetilde{F_1} (r,t)
& =  &
\frac{1}{2 p^3 r^2}{\mathrm e}^{\frac{H t}{2}} \sin \left(p\frac{1-  \e^{-H t}}{H}\right) (p r \cos (p r)-\sin (p r))\,,\\ 
\widetilde{G_0} (r,t)
& =  &
\frac{1}{2 p^2 r}{\mathrm e}^{\frac{H t}{2}} \sin (p r) \sin \left(p\frac{1-  \e^{-H t}}{H}\right).
\end{eqnarray}
Consequently,
\begin{eqnarray} 
\Psi_{dS} (r,\theta ,\phi,t) 
& = &
- {\mathscr{D}}^{co}(t,\partial _t,\partial _x;0) \gamma ^0  \left(
\begin{array}{c}
  \sqrt{\frac{1}{2}}Y_{1}^{-\frac{1}{2}}(\theta ,\phi )\widetilde{F_{1}}  (r,t) \\
 \sqrt{\frac{1}{2}}Y_{1}^{\frac{1}{2}}(\theta ,\phi )\widetilde{F_{1}}  (r,t)  \\
  i\sqrt{\frac{1}{2}} Y_{0}^{-\frac{1}{2}}(\theta ,\phi )  \widetilde{G_0} (r,t)\\
-i\sqrt{\frac{1}{2}}Y_{0}^{\frac{1}{2}}(\theta ,\phi ) \widetilde{G_0} (r,t)  \\
\end{array}
\right)\, \quad \mbox{\rm for }\quad j= \frac{1}{2} \,.
 \end{eqnarray}
Thus, the wave in the de~Sitter space generated by the spherical wave with $M=0$, $\ell=0$, and $j= \frac{1}{2}$ in Minkowski space   is
\begin{equation}
\Psi_{dS} (r, \theta, \phi,t) 
  = 
{\mathrm e}^{ i p\frac{\left(1-e^{-H t} \right)}{H} -\frac{3 H t}{2}}\left(
\begin{array}{c}
-{\mathrm e}^{ -\frac{i \phi }{2} }\frac{\sqrt{\tan \left(\frac{\theta }{2}\right)} (2 \cos (\theta )+1)  }{2 \pi   }\frac{(p r \cos (p r)-\sin (p r))}{p^2 r^2}\\
{\mathrm e}^{\frac{i \phi }{2}}\frac{\sqrt{\cot\left(\frac{\theta }{2}\right)} (2 \cos (\theta )-1)  }{2 \pi   }\frac{(p r \cos (p r)-\sin (p r))}{p^2 r^2}\\
i{\mathrm e}^{ -\frac{i \phi }{2} }\frac{ \sqrt{\tan \left(\frac{\theta }{2}\right)}  }{2 \pi }\frac{\sin (p r)}{ p r}\\
i{\mathrm e}^{ \frac{i \phi }{2}}\frac{ \sqrt{\cot \left(\frac{\theta }{2}\right)}  }{2 \pi  }\frac{\sin (p r)}{ p r}\\
\end{array}
\right)\,. 
\end{equation}
\begin{figure}
\hspace{0.1cm}\scalebox{0.7}{\includegraphics{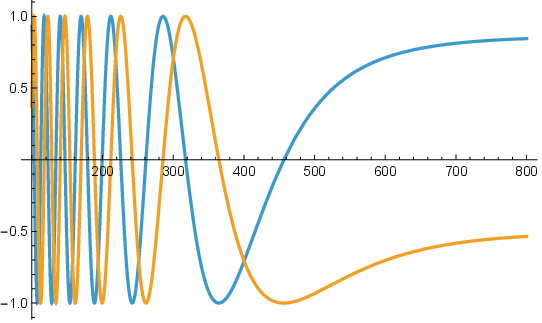}} 
\caption{  $p=1$, $H=0.01$, $ \Im \exp {\frac{i \left(p-p \e^{-H t}\right)}{H}}$ and $\Re \exp {\frac{i \left(p-p \e^{-H t}\right)}{H}} $, $T_{ls.osc} \approx 460$ }
\label{F1}
\end{figure}
The angular part of the solution coincides with one of the initial states. All components of the spinor have the time-dependent factor
$ \exp \left({i p\frac{ \left(1-{\mathrm e}^{-H t}\right)}{H}-\frac{3 H t}{2}}\right)$ indicating the decay  and oscillation in time of the solution. Thus, the best decay rate for such solutions in the $L_\infty$  space is ${\mathrm e}^{-\frac{3 H t}{2}}$ while the cosmological constant causes the damping of oscillations, which vanish after a certain life-span time $T_{ls.osc}$. (See Fig.\ref{F1}.)    The limit as $H$ approaches zero gives a solution in the Minkowski space, where no decay occurs, and the oscillations never vanish, $T_{ls.osc}=\infty$.

 Comparing solutions in these two spaces links time to the Hubble constant and thereby may potentially be used  to measure either $t$ or $H$.

\subsection{Example of particle with imaginary mass $M=  iH$ and $\ell=0$}
\label{SS2.4}

 Of particular interest is the fact that, in the case of the Dirac equation, it is precisely one massless particle and two particles with imaginary mass that obey Huygens' principle, which is one of the most important principles in physics. For this very reason, the present and the next subsections are devoted to a detailed examination of particles with imaginary mass in the expanded universe. A brief and far-from-exhaustive list of works on tachyons in Minkowski space is included in the references (see \cite{Feinberg}-\cite{Moucherek} and references therein).

The next  solution, which obeys the Huygens' principle, we obtain for $M=  iH$, $\ell=0$, and $ j=\ell +\frac{1}{2}$. 
Spherical solution in Minkowski space with $\ell=0$,  
  $M= iH$, $p\geq H$,   is given by (\ref{eelplas1/2}) and (\ref{A4}).
For  $M= iH$, $\ell=0$, and $ j=\ell +\frac{1}{2}$, it follows 
$
 M_+=-\frac{1}{2}H$, and $M_-=\frac{3}{2}H$. Hence, 
\begin{eqnarray}
\label{K1M0_2}
K_1\left(r,t;-\frac{1}{2}H \right)
 :=  
\frac{1}{2} \e^{\frac{H t}{2}},\quad 
K_1\left(r,t;\frac{3}{2}H\right)
 = \frac{1}{4} \e^{-\frac{1}{2} (H t)} \left(\left(1-H^2 r^2\right) \e^{2 H t}+1\right)\,.   
\end{eqnarray}
Further, we use (\ref{eelplas1/2}), (\ref{eelminus1/2}), (\ref{A4}), (\ref{A5}), 
(\ref{Ftilde}), (\ref{Gtilde})  to derive
\begin{eqnarray}
f(r)
= \frac{ \left(p r\cos (p r )-\sin (p r )\right)}{\left(\sqrt{p^2-H^2}+i H\right)  p r^2 },\quad
F_1(r,t) = \cos (p t)\frac{ (p r \cos (p r)-\sin (p r))}{p r^2 \left(\sqrt{p^2-H^2}+i H\right)}\,,
\end{eqnarray}
\begin{eqnarray}
\widetilde{F_1} (r,t)
& =  &
{\mathrm e}^{\frac{H t}{2}} \sin \left(p\frac{1- \e^{-H t}}{H}\right)\frac{\left(\sqrt{p^2-H^2}-i H\right)  (p r \cos (p r)-\sin (p r))}{2 p^4 r^2}\,, 
\end{eqnarray}
\begin{eqnarray}
g(\tau)=\frac{\sin (p \tau )}{p \tau },\quad
G_0 (r,t)
 = 
\cos (p t)\frac{1}{p r}\sin (p r) , 
\end{eqnarray}
\begin{equation}
\widetilde{G_0} (r,t)
  =  
\frac{{\mathrm e}^{\frac{H t}{2}}  \left(\left(H^2 \e^{H t}+p^2\right) \sin \left(p\frac{1-  \e^{-H t}}{H}\right)-H p \left(\e^{H t}-1\right) \cos \left(p\frac{1-  \e^{-H t}}{H}\right)\right)\sin (p r)}{2 p^4 r}\,. 
\end{equation}
The solution 
$\Psi_{dS}(r,\theta,\phi,t) $ is % ={}^{tr}(\Psi_1,\Psi_2,\Psi_3,\Psi_4)
\begin{equation}
\Psi_{dS}(r,\theta,\phi,t)=\left(
\begin{array}{c}
 i {\mathrm e}^{-\frac{ 3  H t}{2}-\frac{ i \phi}{2}  }\frac{\sqrt{\tan \left(\frac{\theta }{2}\right)}(1+2 \cos (\theta ) )}{2 \pi  }R_U(r,t;H,p)\\
  - i {\mathrm e}^{-\frac{3 H t}{2}+\frac{i \phi }{2}}\frac{\sqrt{\cot \left(\frac{\theta }{2}\right)}(2 \cos (\theta )-1) }{2 \pi   }R_U(r,t;H,p)\\
i {\mathrm e}^{-\frac{ 3H t}{2}   -\frac{ i \phi}{2} }\frac{ \sqrt{\tan \left(\frac{\theta }{2}\right)} }{2 \pi   }R_L(r,t;H,p) \\
i {\mathrm e}^{-\frac{3 H t}{2}+\frac{i \phi }{2}}\frac{ \sqrt{\cot  \left(\frac{\theta }{2}\right)}   }{2\pi   } R_L(r,t;H,p) \\
\end{array}
\right)\,,
\end{equation}
where the radial functions $R_U $ and $ R_L$ are
\begin{eqnarray}
R_U(r,t;H,p)
& := &
\frac{(p r \cos (p r)-\sin (p r))}{p^4 r^2}\Bigg[ p \cos \left(p\frac{1-  \e^{-H t}}{H}\right) \left(H  \e^{H t}+i \sqrt{ \left(p^2-H^2\right)}\right)\nonumber\\
&  &
- \left(p^2-i H \sqrt{p^2-H^2} \e^{H t}\right) \sin \left(p\frac{1- \e^{-H t}}{H}\right)\Bigg],\\
R_L(r,t;H,p)
& := &
\frac{\sin (p r)}{p^2 r}\Bigg[  p \cos \left(p\frac{1- \e^{-H t}}{H}\right)+i \sqrt{p^2-H^2} \sin \left(p\frac{1- \e^{-H t}}{H}\right) \Bigg]\,.
\end{eqnarray}

\begin{figure}
\centering
\begin{subfigure}{.5\textwidth}
  \centering
  \includegraphics[width=.8\linewidth]{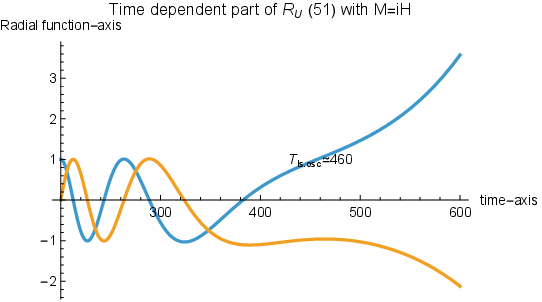}
  %\caption{A subfigure}
  %\label{F2}
\end{subfigure}%
\begin{subfigure}{.5\textwidth}
  \centering
  \includegraphics[width=.8\linewidth]{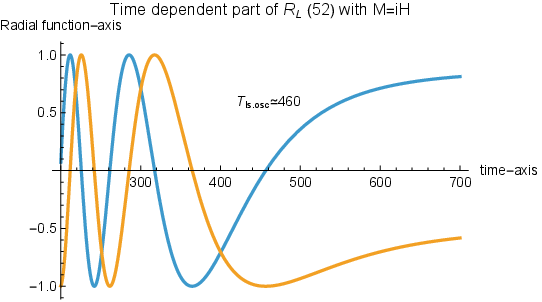}
  %\caption{A subfigure}
  %\label{F2}
\end{subfigure}
\caption{$M=iH$, $p=1$, $H=0.01$, blue is a real part, red is an imaginary part}
\label{F2}
\end{figure}

The factor $\left[  p \cos \left(p\frac{1- e^{-H t}}{H}\right)+i \sqrt{p^2-H^2} \sin \left(p\frac{1- e^{-H t}}{H}\right) \right] $ describes an elliptic harmonic motion in the complex plane (Fig. \ref{F2}).
Thus, the best decay rate for such solutions $\Psi_{dS}$ in the $L_\infty ({\mathbb R}^3)$-space  is  $e^{-\frac{  H t}{2}}$, while the cosmological constant causes the damping of oscillations, which vanish after time $T_{ls.osc}$. (See Fig.\ref{F2} with $T_{ls.osc}\approx 460$.) In the case of Minkowski space,  no decay occurs, and the oscillations never vanish, $T_{ls.osc}=\infty$.

\subsection{Example of of particle with imaginary mass $M= - iH$ and $\ell=0$ } 
 \label{SS2.5}

 The spherical solution in Minkowski space is given by (\ref{eelplas1/2}) and (\ref{A4}). 
Hence, in this case, 
\begin{equation}
f(r)=\frac{\left(\sqrt{p^2-H^2}+i H\right) (p r \cos (p r)-\sin (p r))}{p^3 r^2},\quad g(r) = \frac{\sin (p r)}{p r}\,,
\end{equation}
\begin{equation}
F_1(r,t)\text{:=}\frac{\cos (p t) (p r \cos (p r)-\sin (p r))}{p r^2 \left(\sqrt{p^2-H^2}+i H\right)}\,,
\end{equation}
\begin{eqnarray}
&  &
\widetilde{F_1}(r,t) \\
& := &
\frac{\e^{\frac{H t}{2}} (p r \cos (p r)-\sin (p r)) \left(\left(H^2 \e^{H t}+p^2\right) \sin \big(p\frac{1- \e^{-H t}}{H}\big)-H p \left(\e^{H t}-1\right) \cos \big(p\frac{1- \e^{-H t}}{H}\big)\right)}{2 p^4 r^2  (\sqrt{p^2-H^2}+i H )},\nonumber
\end{eqnarray}
\begin{equation}
G_0(r,t) := \frac{\sin (p r) \cos (p t)}{p r},\quad
\widetilde{G}_0(r,t)=\frac{\e^{\frac{H t}{2}} \sin (p r) \sin \left(p\frac{1- \e^{-H t}}{H}\right)}{2 p^2 r}\,.
\end{equation}
The solution $\Psi_{dS}$  can be written  as follows 
\begin{equation}
\Psi_{dS}(r,\theta,\phi,t)=\left(
\begin{array}{c}
{\mathrm e}^{-\frac{3}{2}  H t-\frac{i \phi }{2}}\frac{\sqrt{\tan \left(\frac{\theta }{2}\right)}(1+2 \cos (\theta ) ) }{4\pi     }   R_U(r,t;H,p) \\
{\mathrm e}^{-\frac{3}{2} H t+\frac{i \phi }{2}}\frac{  \sqrt{\cot \left(\frac{\theta }{2}\right)} (1-2 \cos (\theta ) ) }{4 \pi    } R_U(r,t;H,p) \\
{\mathrm e}^{-\frac{ 3}{2} H t-\frac{i \phi }{2}}\frac{\sqrt{\tan \left(\frac{\theta }{2}\right)} }{4 \pi  } R_L(r,t;H,p) \\
-  {\mathrm e}^{-\frac{3 }{2}H t+\frac{i \phi }{2}}\frac{\sqrt{\cot \left(\frac{\theta }{2}\right)} }{4 \pi   }R_L(r,t;H,p) \\
\end{array}
\right)\,,
\end{equation}
where the radial functions $R_U $ and $ R_L$ are
\begin{eqnarray}
R_U(r,t;H,p)
& := &
\frac{(p r \cos (p r)-\sin (p r))}{p^2 r^2\left(\sqrt{p^2-H^2}+i H\right)}\\
&  &
\times \Big[\left(\sqrt{p^2-H^2}+2 i H\right) \sin \left(p\frac{1-  \e^{-H t}}{H}\right)-i p \cos \left(p\frac{1-  \e^{-H t}}{H}\right)\Big]\,,\nonumber\\
R_L(r,t;H,p)
& := &\frac{i  \sin (p r)}{p^2 r \left(\sqrt{p^2-H^2}+i H\right)} \\
&  &
\times \Bigg[ i \left(H \left(\sqrt{p^2-H^2}+2 i H\right) \e^{H t}+i p^2\right) \sin \left(p\frac{1-  \e^{-H t}}{H}\right) \nonumber\\
&  &
+p \left(H \left(e^{H t}-2\right)+i \sqrt{p^2-H^2}\right) \cos \left(p\frac{1-  \e^{-H t}}{H}\right) \Bigg]\,. \nonumber
\end{eqnarray}
\begin{figure}
\centering
\begin{subfigure}{.5\textwidth}
  \centering
  \includegraphics[width=.8\linewidth]{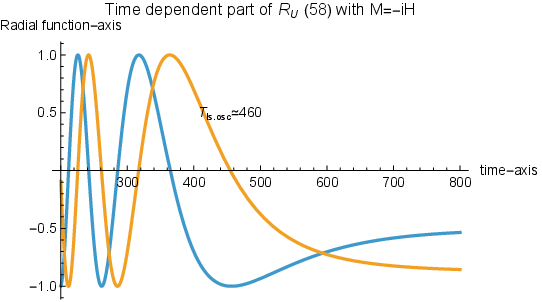}
  %\caption{A subfigure}
  %\label{F3}
\end{subfigure}%
\begin{subfigure}{.5\textwidth}
  \centering
  \includegraphics[width=.8\linewidth]{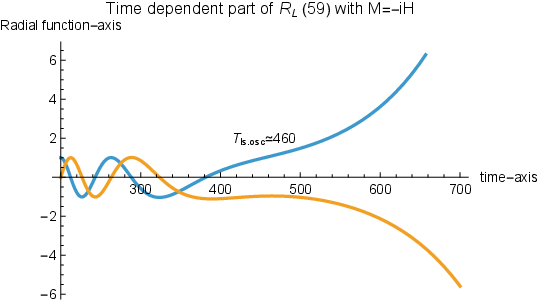}
  %\caption{A subfigure}
  %\label{F3}
\end{subfigure}
\caption{$M=-iH$, $p=1$, $H=0.01$, blue is a real part, red is an imaginary part}
\label{F3}
\end{figure}
Thus, the best decay rate for such solutions $\Psi_{dS}$ in the $L_\infty ({\mathbb R}^3)$  space is $\e^{-\frac{  H t}{2}}$, while the cosmological constant causes the damping of oscillations, which vanish after a life-span time $T_{ls.osc}$ (see Fig.\ref{F3} with $T_{ls.osc}\approx 460$). In the case of Minkowski space   $H=0$, no decay occurs, and the oscillations never vanish, $T_{ls.osc}=\infty$.

\subsection{Solution with an initial wave produced by the hydrogen-like atom}
\label{SS2.6}

For the wave produced by the hydrogen-like atom in Minkowski space,   formulas (\ref{4.1}) and (\ref{4.2}) can also be written in a more compact form due to the factor that is exponentially decaying as $r \to \infty$. 
According to Lemma~2.4~\cite{ArXiV2025}, if
$ F_0(r) \in    C^1  (0,\infty)$, and 
\begin{equation}
\label{intF}
\int_0^\infty  r ^{3/2}|F_0(r)|\,dr<\infty,\quad    F_0(r)=O(r^{-k}), \quad r \to \infty, \quad k> 2,
\end{equation}  
then the solution to the problem
$ 
v_{tt}-  \bigtriangleup v =0$, $v(x,0)=F_0(r)  Y_{\ell m}(\theta, \phi) 
$, $v_t(x,0)=0$, 
is given by
\begin{equation}
 v (r,\theta, \phi,t)  
 =  
Y_{\ell m}(\theta, \phi)\frac{1}{\sqrt{r}}\int_{0}^\infty \left(\int_{0}^\infty      F_0(\rho) J_{\ell+\frac{1}{2}}(\rho \lambda )\rho^{3/2}\, d\rho\right)\cos \left( \lambda t \right) J_{\ell+\frac{1}{2}}(r \lambda )\lambda\, d\lambda\,.
\end{equation} 
For the hydrogen-like atom with (\ref{f_Hyd}) and (\ref{g_Hyd}) as an initial datum, since (\ref{intF}) is fulfilled, the Hankel transform is applicable. Note that the last condition is violated in the case of free-field initial data, which led us to turn to the Riemann function approach. 
Thus, with the aid of the   Hankel transform,  the functions $F_{\ell+1} (r ,t ) $ and $ G_\ell (r ,t ) $ are given by 
\begin{eqnarray}  
F_{\ell+1} (r ,t )  
& = &
\frac{1}{\sqrt{r}}\int_{0}^\infty \left(\int_{0}^\infty      f(\rho) J_{\ell+\frac{3}{2}}(\rho \lambda )\rho^{3/2}\, d\rho\right)\cos \left( \lambda t \right) J_{\ell+\frac{3}{2}}(r \lambda )\lambda\, d\lambda \,,\\ 
G_\ell (r ,t )  
& = &
\frac{1}{\sqrt{r}}\int_{0}^\infty \left(\int_{0}^\infty      g(\rho) J_{\ell+\frac{1}{2}}(\rho \lambda )\rho^{3/2}\, d\rho\right)\cos \left( \lambda t \right) J_{\ell+\frac{1}{2}}(r \lambda )\lambda\, d\lambda
\,,
\end{eqnarray}
where the functions $f $ and $g  $ are given by (\ref{f_Hyd}) amd (\ref{g_Hyd}), respectively. Consequently,
\begin{eqnarray} 
\Psi_{dS} (x,t) 
& = &
{\mathscr{D}}^{co}(t,\partial _t,\partial _x)  i\gamma ^0 \left(
\begin{array}{c}
 \widetilde{F}_{\ell+1} (r,t) i   \sqrt{\frac{\ell -m+\frac{3}{2}}{2 \ell +3}}  Y_{\ell+1}^{m-\frac{1}{2}}(\theta ,\phi ) \\
 \widetilde{F}_{\ell+1} (r,t)i  \sqrt{\frac{\ell +m+\frac{3}{2}}{2 \ell +3}}Y_{\ell+1}^{m+\frac{1}{2}}(\theta ,\phi )  \\
 \widetilde{G_\ell} (r,t) \sqrt{\frac{\ell +m+\frac{1}{2}}{2 \ell +1}}  Y_{\ell }^{m-\frac{1}{2}}(\theta ,\phi ) \\
- \widetilde{G_\ell} (r,t) \sqrt{\frac{\ell -m+\frac{1}{2}}{2 \ell +1}}  Y_{\ell }^{m+\frac{1}{2}}(\theta ,\phi )  \\
\end{array}
\right) \,,
 \end{eqnarray}
where
\begin{eqnarray}
\widetilde{F}_{\ell+1} (r,t)
& = &
 4^{\frac{1}{2}-\frac{M_+}{H}}  \e^{M_+ t}\int_{0}^{\frac{1}{H}\left(1-\e^{-Ht}\right)} 
 \Bigg\{  \left(\left(1+\e^{-H t}\right)^2-(H s)^2\right)^{\frac{M_+}{H}-\frac{1}{2}} \\
&  &
\times  F \left(\frac{1}{2}-\frac{M_+}{H},\frac{1}{2}-\frac{M_+}{H};1;\frac{\left(1-\e^{-H t} \right)^2-(s H)^2}{\left(1+\e^{-H t} \right)^2-(s H)^2}\right) \Bigg\}    \nonumber   \\
&  &
\times \frac{1}{\sqrt{r}}\int_{0}^\infty \left(\int_{0}^\infty      f(\rho) J_{\ell+\frac{3}{2}}(\rho \lambda )\rho^{3/2}\, d\rho\right)\cos \left( \lambda s \right) J_{\ell+\frac{3}{2}}(r \lambda )\lambda\, d\lambda  \, ds\,, \nonumber\\
\widetilde{G_\ell} (r,t)
& := &
 4^{\frac{1}{2}-\frac{M_-}{H}}  \e^{M_- t}\int_{0}^{\frac{1}{H}\left(1-\e^{-Ht}\right)} 
 \Bigg\{  \left(\left(1+\e^{-H t}\right)^2-(H s)^2\right)^{\frac{M_-}{H}-\frac{1}{2}} \\
&  &
\times  F \left(\frac{1}{2}-\frac{M_-}{H},\frac{1}{2}-\frac{M_-}{H};1;\frac{\left(1-\e^{-H t} \right)^2-(s H)^2}{\left(1+\e^{-H t} \right)^2-(s H)^2}\right) \Bigg\}        \nonumber\\
&  &
\times\frac{1}{\sqrt{r}}\int_{0}^\infty \left(\int_{0}^\infty      g(\rho) J_{\ell+\frac{1}{2}}(\rho \lambda )\rho^{3/2}\, d\rho\right)\cos \left( \lambda s \right) J_{\ell+\frac{1}{2}}(r \lambda )\lambda\, d\lambda\, ds\,. \nonumber
\end{eqnarray}

\section{Conclusions} 

The spherical-wave solutions of the Dirac equation are important since they describe relativistic spin-1/2 particles in central potentials and spherically symmetric geometries. 
In astrophysics and cosmology, in curved spacetimes with spherical symmetry, the spherical spinor modes of the Dirac equation are used to describe fermions on black-hole backgrounds, particle creation in cosmological spacetimes, and neutrino propagation in gravitational fields. These  include solutions centered on the  Schwarzschild  and  Reissner-Nordstr\"om black holes \cite{Senjaya_PL}. 
The spherical Dirac waves are central to rigorous spectral analysis to study
  self - adjointness of Dirac operators (see, e.g., \cite{Thaller,Piero} and references therein),  spectrum with Coulomb potentials (see, e.g., \cite{Thaller,Baskin} and references therein),  bound and scattering states (see, e.g., \cite{Kennedy}), and  mixed discrete-continuous spectra. These solutions explain the fine structure of atomic spectra (see, e.g., \cite[Ch.9]{Greiner} and references therein).

Thus, in the present work, the spin-$\frac{1}{2}$ spherical waves in the expanding universe are represented by means of the spinless massless scalar waves (virtual particles) in Minkowski space via the co-factor (diagonalizer) of the Dirac operator and the special integral transform. The integral transform condenses the evolution of those virtual waves in the conformal time from their entry into the expanding universe to the moment of observation.
It is indicated that the best decay rate for massless solutions in the $L_\infty$  space is $\e^{-\frac{3 H t}{2}}$, while the cosmological constant causes the damping of oscillations, which vanish after a certain life-span time $T_{ls.osc}<\infty$. 

\appendix

\section{Appendix. Spherical   Dirac waves in   Minkowski \\spacetime}
\label{ss4}

\setcounter{equation}{0}
\renewcommand{\theequation}{\thesection.\arabic{equation}}

{\noindent
{\bf A.1 Free waves.} Definition of spherical harmonics  has been used here is
\begin{equation} 
 Y_\ell^m(\theta ,\phi )=(-1)^m {\mathrm e}^{i m \phi } \sqrt{\frac{(2 \ell+1) (\ell-m)!}{4 \pi  (\ell+m)!}} P_\ell^m(\cos (\theta ))\,,
\end{equation}
where $P_l^m(\cos (x))$ is the associated Legendre polynomial.
 Then the spherical spin-1/2 wave $\Psi (t,r,\theta ,\phi ) $ in the usual notations (see \cite[Sec.14]{Bethe},\cite[Sec.24]{B-L-P}
) is   
\begin{eqnarray}
\label{eelplas1/2}
\Psi (t,r,\theta ,\phi )
& = &
\left(
\begin{array}{c}
\e^{i t \epsilon } i f(r) \sqrt{\frac{\ell -m+\frac{3}{2}}{2 \ell +3}}  Y_{\ell+1}^{m-\frac{1}{2}}(\theta ,\phi ) \\
 \e^{i t \epsilon } i f(r) \sqrt{\frac{\ell +m+\frac{3}{2}}{2 \ell +3}}Y_{\ell+1}^{m+\frac{1}{2}}(\theta ,\phi )  \\
 \e^{i t \epsilon }g(r) \sqrt{\frac{\ell +m+\frac{1}{2}}{2 \ell +1}}  Y_{\ell }^{m-\frac{1}{2}}(\theta ,\phi ) \\
- \e^{i t \epsilon }g(r) \sqrt{\frac{\ell -m+\frac{1}{2}}{2 \ell +1}}  Y_{\ell }^{m+\frac{1}{2}}(\theta ,\phi )  
\end{array}
\right) , \,\, \mbox{\rm for}\,\, j=\ell+\frac{1}{2},\\
\label{eelminus1/2}
\Psi (t,r,\theta ,\phi )
& = &
\left(
\begin{array}{c}
 \e^{i t \epsilon }i f(r) \sqrt{\frac{\ell +m-\frac{1}{2}}{2 \ell -1}} Y_{\ell-1}^{m-\frac{1}{2}}(\theta ,\phi ) \\
 -\e^{i t \epsilon }i f(r) \sqrt{\frac{\ell -m-\frac{1}{2}}{2 \ell -1}}  Y_{\ell-1}^{m+\frac{1}{2}}(\theta ,\phi )\\
\e^{i t \epsilon } g(r) \sqrt{\frac{\ell -m+\frac{1}{2}}{2 \ell +1}} Y_{\ell}^{m-\frac{1}{2}}(\theta ,\phi ) \\
\e^{i t \epsilon }  g(r) \sqrt{\frac{\ell +m+\frac{1}{2}}{2 \ell +1}} Y_{\ell}^{m+\frac{1}{2}}(\theta ,\phi ) 
\end{array}
\right) , \,\, \mbox{\rm for}\,\, j=\ell-\frac{1}{2}.
\end{eqnarray} 
  The radial functions  $f(r) $ and $g(r) $   are (see  \cite[Sec.24]{B-L-P})
\begin{eqnarray}
\label{A4}
f(r)=-c_1\frac{ 1  }{M+\epsilon }\sqrt{\frac{\pi p}{2r}} J_{\ell+\frac{3}{2}}(p r), \quad
g(r) =c_1 \sqrt{\frac{\pi }{2 p r}} J_{\ell+\frac{1}{2}}(p r)\,,\quad \mbox{\rm for}\quad j=\ell+\frac{1}{2}\,,\\
\label{A5}
 f(r)
  =  
c_1\frac{1}{M+\epsilon }\sqrt{\frac{\pi p}{2r}} J_{\ell-\frac{1}{2}}(p r),\quad 
  g(r)
  =  
c_1 \sqrt{\frac{\pi}{2pr}} J_{\ell+\frac{1}{2}}(p r)\,,\quad \mbox{\rm for}\quad j=\ell-\frac{1}{2}\,,
\end{eqnarray}
where $ \ell$ is orbital angular momentum, $j$ is total angular momentum, $ \epsilon$ is energy, and $ M^2+p^2 = \epsilon ^2$.

{\noindent
{\bf A.2 Hydrogen-like atom wave function.} 
The following explicit expressions for the
normalized radial Dirac eigenfunctions (\cite[(14.37)]{Bethe} and \cite[(1.5.20)]{Akhiezer}) are 
\begin{eqnarray}
\label{f_Hyd}
f(r)
& = &
-C_-(\gamma,n',N,a_0,Z,\kappa)\,\mbox{\rm e}^{-  Z_N r}  r  ^{\gamma -1}\\
&  &
\times \left(n'F\left(-n'+1,2 \gamma +1,2Z_N r \right)+(N-\kappa )F\left(-n',2 \gamma +1,2Z_Nr \right)\right)\,,  \nonumber\\
\label{g_Hyd}
g(r)
& = &
-C_+(\gamma,n',N,a_0,Z,\kappa)\,\mbox{\rm e}^{-Z_Nr }  r  ^{\gamma -1}\\
&  &
\times\left(-n' F\left(-n'+1,2 \gamma +1,2 Z_Nr\right)+(N-\kappa )F\left(-n',2 \gamma +1,2 Z_Nr\right)\right) \,, \nonumber
\end{eqnarray}
where   $F (\alpha, \beta,z)$ is the  confluent  hypergeometric function and  
the numbers   $C_\pm(\gamma,n',N,a_0,$ $Z,\kappa) $, $Z_N:=\frac{ Z}{Na_0} $ are defined by the physical characteristics of the state of atom:
\begin{equation}
C_{\pm}(\gamma,n',N,a_0,Z,\kappa)
 = 
\frac{\sqrt{\Gamma \left( 2 \gamma +n'+1\right)}}{\Gamma (2 \gamma +1)\sqrt{n'!}}\frac{\sqrt{ 1\pm \epsilon }}{\sqrt{4 N (N-\kappa )}}\left( 2 Z _N \right)^{3/2+\gamma -1}\,,
\end{equation}
(for details, see \cite[(14.9),(14.16),(14.22),(14.24),(14.28),(14.29)]{Bethe} and \cite[(19a)]{Greiner}).

\end{document}